# Scheduling with regular performance measures and optional job rejection on a single machine

Baruch Mor[1], Dana Shapira[2]

[1]Department of Economics and Business Administration, Ariel University, Israel

[2]Department of Computer Science, Ariel University, Israel

Corresponding Author:

Baruch Mor,

Department of Economics and Business Administration,

Ariel University,

Ariel, 40700 Israel.

Phone: + 972 39765 711

Fax: + 972 3 9066 658

Email: baruchm@ariel.ac.il

**Abstract**

We address single machine problems with optional *job–rejection,* studied recently in Zhang et al. [21] and Cao et al. [2]. In these papers, the authors focus on minimizing *regular performance measures*, i.e., functions that are non-decreasing in the jobs completion time, subject to the constraint that the total rejection cost cannot exceed a predefined upper bound. They also prove that the considered problems are ordinary NP-hard and provide pseudo-polynomial-time *Dynamic Programming* (DP) solutions. In this paper, we focus on three of these problems: makespan with release-dates; total completion times; and total weighted completion, and present enhanced DP solutions demonstrating both theoretical and practical improvements. Moreover, we provide extensive numerical studies verifying their efficiency.

## 1. Introduction

During the last decade, scheduling problems with optional job-rejection have attracted considerable interest from researchers. Unlike the traditional approach in deterministic scheduling theory, in which all jobs must be processed, in the considered family of problems we are given the choice of job rejection. In a recent and a comprehensive survey, Shabtay et al. [20] state that " … in many practical cases, mostly in highly loaded make-to-order production systems, accepting all jobs may cause a delay in the completion of orders which in turn may lead to high inventory and tardiness costs." Thus, the option of job-rejection provides the production manager with a new level of freedom, and allows him to accept (process) only a subset of the given jobs, while rejecting (out-sourcing or refusing to process) the complementary subset. Unquestionably, the production manager must consider the cost accompanied by the rejection, as each rejected job incurs a job-dependent penalty.

The survey of Shabtay et al. [20] also presents both the theoretical and practical significance of enabling job-rejection in scheduling problems and describes an abundance of problems studied in this area in the past few years. Studies published later include e.g., Domaniç and Plaxton [3] consider scheduling unit jobs with a common deadline to minimize the sum of weighted completion times and rejection penalties. Shabtay [19], studies the single machine serial batch scheduling problem with rejection to minimize total completion time and total rejection cost. Li and Zhao [14] focus on deteriorating jobs scheduling on a single machine with release dates, rejection and a fixed non-availability interval. Different parallel machine settings with job-rejection were considered in several studies: Ou et al. [17] work on an improved heuristic; Jiang and Tan [12] consider non-simultaneous machine available time; Ma and Yuan [15] examine online scheduling with rejection to minimize the total weighted completion time. He et al. [9, 10] investigate scheduling a single machine with parallel batching to minimize makespan and total rejection cost, and improve algorithms for single machine scheduling with release dates and rejections, respectively. Ou et al. [16] present faster algorithms for single machine scheduling with release dates and rejection. Wang et al. [25] study bi-criteria scheduling problems involving job rejection, controllable processing times and rate-modifying activity. Zhang et al. [22] provide approximation algorithms for precedence-constrained identical machine scheduling with rejection. Zou and Miao [24] focus on the single machine serial batch scheduling problems with rejection. Li and Chen [13] research scheduling with rejection and a deteriorating maintenance activity on a single machine. Agnetis and Mosheiov [1] consider scheduling with job-rejection and position-dependent processing times on proportionate flowshops. Gerstl and Mosheiov [5] propose DPs and heuristics for single machine scheduling problems with generalized due-dates and job-rejection. Gerstl et al. [6] explore minmax scheduling problems with acceptable lead-times, and propose extensions to position-dependent processing times, due-window and job-rejection. Zhong et al. [23] address scheduling with release times and rejection on two parallel machines.

In this paper, we address several fundamental single-machine problems with optional *job-rejection,* studied in Zhang et al. [21] and Cao et al. [2]. In these studies, the authors focus on minimizing *regular performance measures*, i.e., functions that are non-decreasing in the jobs' completion time, subject to the constraint that the total rejection cost cannot exceed a predefined upper bound. The authors prove that the considered problems are ordinary NP-hard and provide DP algorithms for all problems, except for the total completion time criteria. Zhang et al. [21] and Cao et al. [2] present FPTAS for minimizing the makespan with a release-date problem and minimizing the total weighted completion time problem, respectively. In this paper, we focus on three of these problems: makespan with release-dates, total completion times, and total weighted completion. We provide DP solutions demonstrating both theoretical and practical improvements. Moreover, we have conducted extensive numerical studies to all solutions, which empirically validates the capability of our DPs to solve large-size instances for the first two problems. For the third problem, the DP is shown to be suited for medium sized orders.

Our paper is constructed as follows. Section 2 provides the formulation of the general problem. In Sections 3, 4 and 5 we present the DP algorithms for the problems: minimizing the makespan with release-dates, minimizing the total completion time, and minimizing the total weighted completion time, respectively. Section 6 concludes.

## 2. Notation and formulation

A set $J$ of $n$ jobs needs to be processed on a single machine. All the jobs are available at time zero, and preemption is not allowed. The scheduler is given the option to accept (process) a subset $A$ of the jobs and to reject the complementary set, $R$, thus $J = A \cup R$ and $A \cap R = \emptyset$. The processing time of job $j$ is denoted by $p_j$, $j = 1, \ldots, n$. For a given schedule, $C_j$ denotes the completion time of job $j, j \in A$. The rejection cost of job $j$ is denoted by $e_j$, $j = 1, \ldots, n$ and the total accepted rejection upper bound is denoted by $U$.

In all three problems discussed in these paper the goal is to find an optimal schedule that minimizes a non-decreasing function, of the completion times of the accepted jobs, subject to a constraint on the total rejection cost $U$.

For the first problem, we add a job dependent release-date denoted by $r_j$, $j \in J$ and the scheduling criteria is the makespan. Thus, the scheduler needs to minimize the makespan of the accepted jobs, $C_{max} = \sum_{j \in A} p_j$, subject to the constraint that total rejection cost does not exceed the upper bound $U$. Using the three-field notation, introduced by Graham et al. [8], the first problem denoted by **P1** is:

**P1**: $1/r_j, \sum_{j \in R} e_j \leq U/C_{max}$.

In our second problem, we aim to minimize the total completion time given that the total rejection cost cannot exceed $U$, thus the problem denoted by **P2** is:

**P2**: $1/r_j, \sum_{j \in R} e_j \leq U/\sum C_j$.

The last problem addressed in this paper is minimizing the total weighted completion time subject to the constraint that total rejection cost does not exceed the upper bound $U$. Let $w_j$ denote the job dependent weight of job $j \in J$ and the third problem denoted by **P3** is:

**P3**: $1/r_j, \sum_{j \in R} e_j \leq U/\sum w_j C_j$.

## 3. Problem $P1$: $1/r_j, \sum_{j \in R} e_j \leq U/C_{max}$

Zhang et al. [21] prove that the problem $1/\sum_{j \in R} e_j \leq U/C_{max}$, is NP-hard based on the reduction from the well-known *minimization Knapsack* problem (see Güntzer and Jungnickel [7]). Thus, the extended problem, $P1$, in which release dates of the jobs are introduced, is also NP-hard. The authors then provide an $O\left(n\left(r_n + \sum_{j=1}^{n} p_j\right)\right)$ time DP algorithm for solving the problem, proving that the problem is ordinary NP-hard. In this section, we provide an improved DP algorithm for Problem $P1$, with running time of $O(nU)$. Based on the theorem by Jackson [11] who showed that the problem $1/r_j/C_{max}$ is optimally solved by sorting the jobs in a non-decreasing order of the release dates, i.e., $r_1 \leq r_2 \leq \cdots \leq r_{n-1} \leq r_n$, the jobs are first sorted accordingly. We define the following state variables:

Let $f(j, i)$ denote the minimum completion time for the partial schedule of jobs $1, \ldots, j$ with maximum rejection cost $i$. At each iteration of the DP, the minimum completion time of jobs 1 to $j$, having an upper bound $i$ on the rejection cost is computed, based on the completion time of jobs 1 to $j-1$, with an upper bound rejection cost of either $i$ or $i - e_j$. At each stage, one needs to decide whether to accept or reject job $j$, as follows,

   i. Job $j$ *must* be accepted in case its rejection cost exceeds the current rejection limit $i$.

   ii. Job $j$ *may* be accepted in case its release time is at most or higher than the minimum completion time of jobs 1 to $j - 1$. In these cases, upon acceptance, the completion time should be increased by $p_j$ or by $r_j + p_j$, in correspondence.

   iii. Job $j$ *may* be rejected in case it minimizes the total completion time.

Thus, we obtain the following recursion formula,

*Dynamic programming algorithm DP1:*

$$f(j,i) = \begin{cases} \begin{cases} \min(f(j-1,i-e_j), r_j + p_j), & f(j-1,i) < r_j \\ \min(f(j-1,i-e_j), f(j-1,i) + p_j), & f(j-1,i) \geq r_j \end{cases}, & e_j \leq i \\ \begin{cases} r_j + p_j, & f(j-1,i) < r_j \\ f(j-1,i) + p_j, & f(j-1,i) \geq r_j \end{cases}, & e_j > i \end{cases}. \quad (1)$$

The boundary conditions are:

$$f(j, 0) = \begin{cases} C_{j-1} + p_j, & C_{j-1} \geq r_j \\ r_j + p_j, & C_{j-1} < r_j \end{cases}, \forall\, 0 \leq j \leq n, \text{ as } i = 0 \text{ implies that no job can be rejected;}$$

$f(0, i) = 0, \ \forall\, 0 \leq i \leq \min(\sum_{j=1}^n e_j, U)$, implies that no jobs are considered.

The optimal solution is given by $f(n, U)$.

*Theorem 1:* The computational complexity of $DP1$ is $O(n \min(\sum_{j=1}^n e_j, U))$.

*Proof:*

Using the formula in (1), the dynamic programming is calculated for every job $j$, $1 \leq j \leq n$, and every rejection cost $i \leq \min(\sum_{j=1}^n e_j, U)$, thus implying a $O(n \min(\sum_{j=1}^n e_j, U))$ processing time. Reconstructing the solution is done by backtracking, starting at $f(n, \min(\sum_{j=1}^n e_j, U))$ and ending at $f(0,0)$, for an addition of $O(n + \min(\sum_{j=1}^n e_j, U))$ operations, for a total of $O(n \min(\sum_{j=1}^n e_j, U))$ processing time.

*Example 1:*

Consider the following instance of the problem with $n = 10$ and $U = 93$, and the jobs are sorted in non-decreasing order of the release dates and renumbered.

The job processing times are,

$p = (47, 41, 20, 42, 31, 15, 12, 21, 18, 24)$.

The job release dates are,

$r = (18, 70, 81, 102, 144, 302, 316, 354, 359, 365)$.

The job rejection costs are,

$e = (44, 14, 20, 28, 16, 29, 46, 32, 38, 1)$.

Applying $DP1$, we obtain the following optimal solution:

The set of rejected jobs is, $R = (J_5, J_8, J_9, J_{10})$.

The set of accepted jobs is, $A = (J_1, J_2, J_3, J_4, J_6, J_7)$.

The optimal makespan is $C_{max} = 329$, and the solution valid since $\sum_{j \in R} e_j = 87 \leq 93 = U$.

*Numerical study:*

We performed numerical tests in order to measure running times of $DP1$. We coded all the experiments in this paper in C++ and executed them on an Intel (R) Core ™ i5-6200U CPU @ 2.30 GHz 4.0 GB RAM platform. We generated random instances having $n = 500, 1000, 1500$ and 2000 jobs. The job processing times and the job rejection costs were generated uniformly in the interval $[1, 50]$. Let $p_{max} = \max\{p_j | j \in \mathcal{J}\}$ denote the maximal processing time and $e_{max} = \max\{e_j | j \in \mathcal{J}\}$ denote the latest release date, thus $p_{max} = e_{max} = 50$. The release-dates were generated uniformly in the interval $[0, 0.80np_{max}]$ to reflect full spectrum of zero to nearly the total sum of processing times. To avoid trivial solutions, i.e. all the jobs are either rejected or accepted, the total rejection cost upper bound ($U$), was generated uniformly in the interval $[0.20nr_{max}, 0.30nr_{max}]$. Actually, the generated $U$-values guarantee approximately equal number of rejected and accepted jobs. For each set of $n$ and $U$, 20 instances were constructed and solved. Table 1 presents the average and worst case running times in milliseconds. The number of jobs, $n$, is given in the first column. The considered intervals in which $U$ was chosen uniformly, is given in the second column. The third and fourth columns present the average and worst case running times, respectively. The results indicate that $DP1$ is extremely efficient and can solve large-size problems. In particular, the worst-case running time for problems of 2000 jobs did not exceed 0.58 seconds.

| $n$ | $U$ | **Average running time [msec]** | **Worst case running time [msec]** |
|---|---|---|---|
| 500 | [5000, 7500] | 30.106 | 43.731 |
| 1000 | [10000, 15000] | 116.391 | 146.718 |
| 1500 | [15000, 22500] | 276.182 | 327.741 |
| 2000 | [20000, 30000] | 484.219 | 575.610 |

*Table 1: Average and worst case running times of $DP1$ algorithm for Problem **P1**.*

## 4. Problem P2: $1/\sum_{j\in R} e_j \leq U/\sum C_j$

Zhang et al. [21] proved that the problem **P2**, is NP-hard, based on reduction from the *even–odd partition* problem (Garey and Johnson [4]), but did not present a solution procedure for the problem. In the following, we provide a DP algorithm for Problem **P2** and thus prove that it is ordinary NP-hard. It is well-known that the Shortest Processing Time first (SPT) rule, i.e., $p_1 \leq p_2 \leq \cdots \leq p_n$ is optimal for $1//\sum C_j$, (see Pinedo [18]). Thus, we start the DP by sorting the jobs in SPT order.

Let $f(j, i)$ denote the total completion time for the partial schedule of jobs $1, \ldots, j$ and maximum rejection cost $i$. Similar to $DP1$, at each iteration of the DP, one needs to decide whether to accept job $j$, and thus to increase the total completion time, or rather to reject job $j$, in case its rejection cost does not exceed the current rejection cost $i$.

Thus, the formal recursion function is,

*Dynamic programming algorithm DP2:*

$$f(j,i) = \begin{cases} \min\big(f(j-1,i) + C_j, f(j-1, i - e_j)\big), & e_j \leq i \\ f(j-1, i) + C_j, & e_j > i \end{cases} \quad (2)$$

The boundary conditions are:

$f(j, 0) = \sum_{k=1}^{j} C_k, \quad \forall\, 0 \leq j \leq n.$

$f(0, i) = 0, \quad \forall\, 0 \leq i \leq \min(\sum_{j=1}^{n} e_j,\ U).$

The optimal solution is given by $f(n, U)$.

*Theorem 2: The computational complexity of DP2 is $O\big(n \min(\sum_{j=1}^{n} e_j,\ U)\big)$.*

*Proof*: See the proof for Theorem 1 in Section 3.

*Example 2:*

Consider the following instance of the problem with $n = 10$ and $U = 66$, and the jobs are sorted in SPT order and renumbered.

The job processing times are,

$p = (15, 18, 23, 24, 28, 33, 36, 38, 46, 47).$

The job rejection costs are,

$e = (21, 46, 7, 10, 15, 32, 33, 10, 46, 29)$.

Applying $DP2$, we obtain the following optimal solution:

The set of rejected jobs is, $R = (J_1, J_3, J_4, J_5, J_8)$.

The set of accepted jobs is, $A = (J_2, J_6, J_7, J_9, J_{10})$.

The optimal makespan is $\sum C_j = 469$, and the solution valid since $\sum_{j \in R} e_j = 63 \leq 66 = U$.

*Numerical study:*

We adapted the scheme planned for Problem **P1** to suit Problem **P2** by eliminating the jobs release-dates. Again, to avoid trivial solutions the total rejection cost upper bound ($U$), was generated uniformly in the interval $[0.10 nr_{max}, 0.15 nr_{max}]$. Table 2, having the same structure as Table 1, presents the average and worst case running times. The worst-case running time for instances of 2000 jobs did not exceed 193 msec, demonstrating that $DP2$ is extremely efficient and can be used to solve real life-size problems.

| $n$ | $U$ | Average running time [msec] | Worst case running time [msec] |
|---|---|---|---|
| 500 | $[2500, 3750]$ | 10.712 | 13.159 |
| 1000 | $[5000, 7500]$ | 40.892 | 50.716 |
| 1500 | $[7500, 11250]$ | 83.416 | 118.952 |
| 2000 | $[10000, 15000]$ | 151.841 | 192.681 |

*Table 2: Average and worst case running times of $DP2$ algorithm for Problem **P2**.*

## 5. Problem *P3*: $1/\sum_{j\in R} e_j \leq U / \sum w_j C_j$

In this section we address Problem **P3**, studied in Cao et al. [2]. The authors prove that the considered problem is NP-hard, and present a DP solution with computational complexity of $O(n^3 p_{max}^2 w_{max})$ time, where $p_{max} = \max\{p_j|\ j \in \mathcal{J}\}$ and $w_{max} = \max\{w_j|\ j \in \mathcal{J}\}$. Therefore, they suggest an FPTAS algorithm. The suggested DP runs in $O(n \cdot \sum_{j=1}^{n} p_j \cdot U)$. As $\sum_{j=1}^{n} p_j \leq n \cdot p_{max}$, our solution is at least faster by a factor of $n$. We start our DP by sorting the jobs in WSPT (Weighted Shortest Processing Time) first order, i.e., non – increasing order of $\frac{p_1}{w_1} \leq \frac{p_2}{w_2} \leq \cdots \leq \frac{p_n}{w_n}$.

Let $f(j, t, i)$ denote the total weighted completion time for the partial schedule of jobs $1, \ldots, j$, having completion time $t$ and maximum rejection cost $i$. Similar to $DP1$, at each iteration of the DP, one needs to decide whether to accept job $j$, and thus increase the total weighted completion time, or rather reject job $j$, in case job $j$'s rejection cost does not exceed the current rejection limit $i$. The formal recursion function is as follows.

*Dynamic programming algorithm DP3*

$$f(x) = \begin{cases} \infty, & p_j > t \text{ and } e_j > i \\ f(j-1, t-p_j, i) + w_j t, & p_j \leq t \text{ and } e_j > i \\ f(j-1, t, i-e_j), & p_j > t \text{ and } e_j \leq i \\ \min\left(f(j-1, t-p_j, i) + w_j t, f(j-1, t, i-e_j)\right), & p_j \leq t \text{ and } e_j \leq i \end{cases} \quad (3)$$

The boundary conditions are:

$f(0,0,i) = 0$ for all $i \leq U$, as if the set of jobs is empty, their total completion time is 0, and so is their cost.

$f(0, t, i) = \infty$ if $t \neq 0$ and $i \leq U$, as if the set of jobs is empty, their total completion time must be 0.

$f(j, t, 0) = \sum_{k=1}^{j} w_k C_k$, and $t = \sum_{k=1}^{j} p_k$. In case the rejection upper bound is 0, all jobs must be processed.

The optimal solution is given by $\min\{f(n, t, i) | 0 \leq t \leq \sum_{j=1}^{n} p_j, 0 \leq i \leq \sum_{j \in \mathcal{J}} e_j\}$.

*Theorem 3: The computational complexity of DP3 is $O\left(n \cdot \sum_{j=1}^{n} p_j \cdot \min\left(\sum_{j=1}^{n} e_j, U\right)\right)$.*

The proof is similar to that of Theorem 1 in Section 3.

At first sight, one could have used formula (2) and replaced the additive $C_j$ with $w_j C_j$. However, the following example intuitively explains the main difference between $DP2$ and $DP3$. Consider for example 5 jobs having processing times 24, 44, 34, 25 and 47, rejection costs 19, 19, 36, 40 and 34, job weights 16, 15, 11, 8 and 5, and rejection upper bound $U = 55$, respectively. The jobs $\frac{p_j}{w_j}$ are 1.5, 2.9, 3.1, 3.1 and 9.4, thus the jobs are given in WSPT order. When job $J_1$ is chosen, its charge is $w_1 p_1 = 16 * 24 = 384$. Alternatively, if Job $J_3$ was the first accepted job, it would have incurred a cost of $w_3 p_3 = 11 * 34 = 374$, which is less than the cost of job $J_1$. At a superficial glance it seems cheaper to process job $J_3$ rather than job $J_1$. However when considering the succeeding jobs and taking into account their relative high rejection costs, 40 and 34, it is better to process jobs $J_4$ and $J_5$, than reject them. Thus, as the processing time of job $J_1$ is less than the processing time of job $J_3$, adding the costs of jobs $J_4$ and $J_5$ results in:

$w_1 p_1 + w_4(p_1 + p_4) + w_5(p_1 + p_4 + p_5) < w_3 p_3 + w_4(p_3 + p_4) + w_5(p_3 + p_4 + p_5)$, which is not as expected.

The fact that jobs may be rejected necessitates the computation of the cost of each job at every time unit, as if it was its completion time, which justifies the addition of another state variable $t$.

*Example 3:*

Consider the following instance of the problem with $n = 10, U = 88$, and the jobs are renumbered according to their WLPT order.

The job processing times are,

$p = (5, 20, 28, 13, 33, 35, 16, 35, 41, 48)$.

The job weights are,

$w = (8, 20, 24, 9, 18, 18, 8, 17, 19, 1)$.

The job rejection costs are,

$e = (36, 23, 6, 31, 3, 40, 22, 10, 32, 21)$.

Applying $DP3$, we obtain the following optimal solution,

The set of rejected jobs is, $R = (J_2, J_3, J_5, J_8, J_9)$,

The set of rejected jobs is, $A = (J_1, J_4, J_6, J_7, J_{10})$, $\sum w_j C_j = 1825$, and $\sum_{j \in R} e_j = 74 \leq 88 = U$

*Numerical study:*

We adapted the scheme planned for Problem $P2$ to suit Problem $P3$ by adding *job-dependent* weights, which were generated uniformly in the interval $[1, 25]$. To guarantee solutions with approximately equal numbers of rejected and accepted jobs the total rejection cost upper bound ($U$), was generated uniformly in the interval $[0.15 nr_{max}, 0.20 nr_{max}]$. Table 3, presents the average and worst case running times in the same format as Table 1. As can be seen, the empirical experiment results confirm that $DP3$ is efficient and can be used to solve medium size instances, the worst-case running time for problems of 40 jobs did not exceed 93 msec.

| $n$ | $U$ | Average running time [msec] | Worst case running time [msec] |
| --- | --- | --- | --- |
| 10 | $[75, 100]$ | 1.235 | 2.878 |
| 20 | $[150, 200]$ | 7.584 | 9.862 |
| 30 | $[225, 300]$ | 26.876 | 38.351 |
| 40 | $[300, 400]$ | 64.850 | 92.279 |

*Table 3: Average and worst case running times of $DP3$ algorithm for Problem $P3$*

## 6. Conclusions

In this paper, we focused on minimizing regular performance measures with optional job-rejection. We introduced improved time complexity DP algorithms to well-known DPs in scheduling theory. Moreover, extensive numerical studies were presented, that support our achievements, and suggest that the enhanced DPs, for minimizing the makespan and minimizing the total completion time, are extremely fast and efficient, and thus suitable for solving large size real life problems.